# A NOVEL HASH BASED LEAST SIGNIFICANT BIT (2-3-3) IMAGE STEGANOGRAPHY IN SPATIAL DOMAIN

G.R.Manjula[1] and AjitDanti[2]

[1]Department of CS&E, JNN College of Engg.,Shimoga-577204, India
[2]Department of MCA, JNN College of Engg., Shimoga-577204, India

## *ABSTRACT*

*This paper presents a novel 2-3-3 LSB insertion method. The image steganography takes the advantage of human eye limitation. It uses color image as cover media for embedding secret message.The important quality of a steganographic system is to be less distortive while increasing the size of the secret message. In this paper a method is proposed to embed a color secret image into a color cover image. A 2-3-3 LSB insertion method has been used for image steganography. Experimental results show an improvement in the Mean squared error (MSE) and Peak Signal to Noise Ratio (PSNR) values of the proposed technique over the base technique of hash based 3-3-2 LSB insertion.*

## 1. INTRODUCTION

Now a days there is a rapid development of the Internet and telecommunication techniques. Importance of information security is increasing. An application such as secret communication, copyright protection, etc, increases the need for research of information hiding systems. Cryptography and Steganography are the major areas which work on Information Hiding and Security.

Steganography is a process of hiding information. It conceals that the communication is taking place therefore when using steganography there is always secret information is being transmitted and we try to make this information not to be discovered just by the intended receiver. The sender hides a message into a cover file likes for e.g. (image, audio,video) and tries to conceal the existence of that message, later the receiver gets this cover file and detects the secret message and receives it.

Steganography means, cover writing it's origin is old and backs to Golden age of Greece when people at that time had different practices to hide writing for e.g. writing on a wooden tablet and then covering it by wax, making a tattoo on a messenger head after shaving his hair and let his hair grows up again and then send him to the receiver where his hair was shaved there again to get the message. Other steganography techniques like using invisible ink for writing between lines, microdots and using character arrangement are also used [1][2][3][4].

Digital steganography has many applications in our life. When sensitive data is transmitted from one place to another they have to be protected from modifying, copying and claiming their ownership. There must be a way to provide availability, integrity, confidentiality services to the information exchanged. Steganography will provide these services [5][6][7].

Figure 1 shows the steganographic system. The object which is used to hide secret information is called cover object. Stego message is referred as a message that is obtained byembedding secret message into cover message. The hidden information may be either plain text, or images etc.

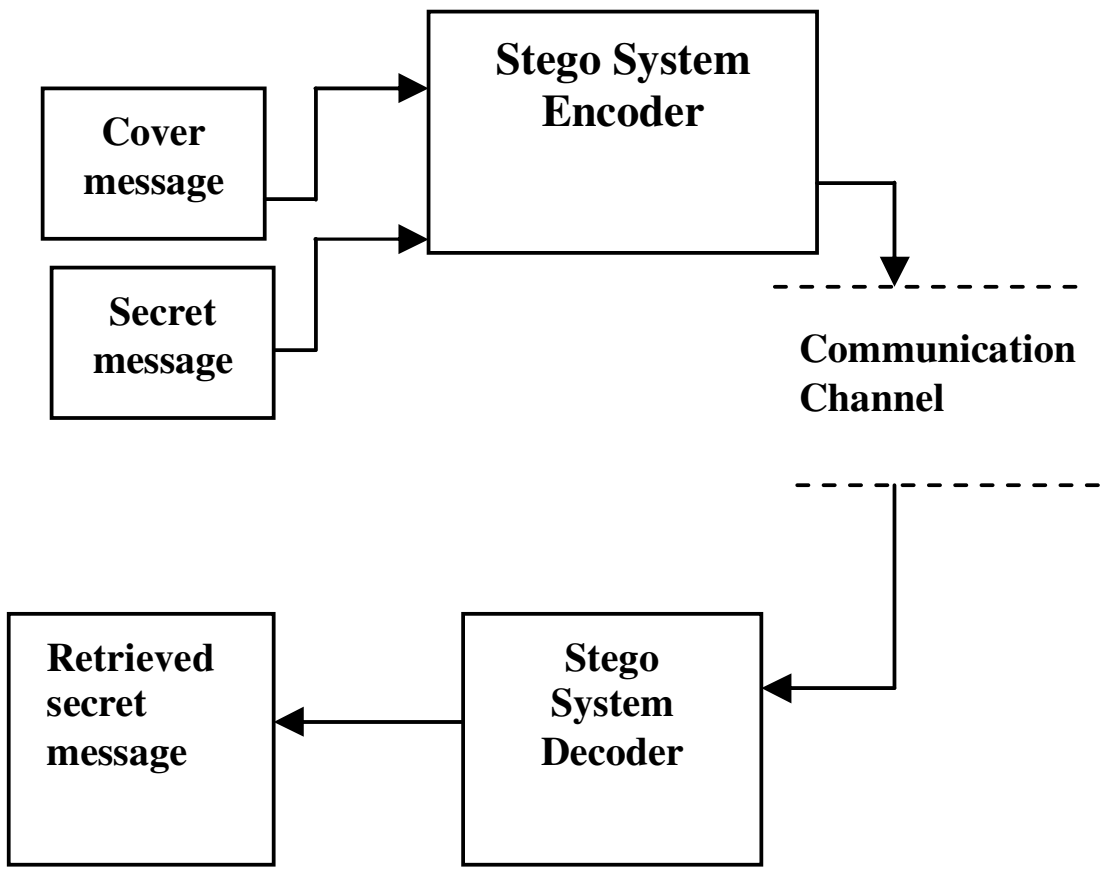

Figure 1:BasicSteganographic system

In image based steganography, it is desirable that a steganography technique is able to hide as many secret message bits as possible in an image in such away it will not affect the most two important requirements that are essential for hiding process and researchers take care about[8][9]:

1. Imperceptibility/security: which means that human eye cannot distinguish between the original image: (the image before hiding process) and the stego- image (the image after hiding process), in other words the hiding process cannot be detected.
2. Capacity: means amount of secret data that can be inserted in a cover media.

The relationship between the above two requirements should be balanced, for e.g. if we increase the capacity more than a specified threshold value then the Imperceptibility will be affected and so on, therefore the parameters of digital steganography technique should be chosen very carefully.Application of Steganography varies from military, industrial applications to copyright and Intellectual Property Rights (IPR). By using lossless steganographic systems secret messages can be sent and received securely [2].

In this paper a hash based LSB 2-3-3 Technique is proposed in spatial domain. An application of the algorithm is illustrated with Color Image file as a cover medium. The results obtained are good and encouraging, compared with based LSB 3-3-2 technique. The rest of the paper is arranged as follows, section 2 does Literature survey of the recent steganographic techniques. In section 3 the proposed hash based LSB 2-3-3 Technique has been described. The algorithm is explained in section 4. Section 5 gives detailed results and performance evaluation of proposed 2-3-3 with 3-3-2 technique Conclusion is presented in Section 6.

## 2. LITERATURE SURVEY

Some of the methods of hiding data in a secure manner are discussed below.

Arvind Kumar and Km Pooja's paper on Steganography [1] describes Steganography as a useful tool that allows covert transmission of information over the communications channel. The performances of some of the steganography tools are analyzed. Combining secret image with the cover image gives the stego image. The hidden image is difficult to detect without proper knowledge embedding method. They also compare between Cryptography and Steganography. Various Steganography software applications are also discussed.

The following table summarizes the work done in the field of image based steganography.

| SL No | Techniques | Methods | Benefits | Drawbacks | Reference No |
|---|---|---|---|---|---|
| 1 | 1.Triple A<br>2.Pixel Indicator | 1. Takes the message, the carrier image, and the password based generated key as inputs and produces the message hidden inside the carrier image.<br>2. In RGB, selects one as indicator channel and the other two to hide data | 1. increased the capacity ratio and the security level of the concealment<br>2. randomization of indicator channel | 1. Key management overhead<br>2. Capacity may reduce | [2] |
| 2 | Truth Table | 1.truth table based on RGB indicator pixel<br>2.additional modulo<br>3.pixel indicator | The transfer medium is made obscure | It is too complex to understand | [3] |
| SL NO | Technique | Method | Benefit | Drawback | Reference No |
| 3 | Hiding Capacity based on RGB images | Data is encrypted using secret key | One third of cover image is saved | It employs secret key | [4] |
| 4 | Hiding secret | Secret data are | Randomness of | Index | |
| | data using LSB technique | segmented into Even segment and Odd segment. Based on this bits are hidden | the index channel | channel is not used efficiently | [5] |
| 5 | LSB with private stego-key | embeds binary bit stream in 24-bits color image (Blue channel) or in 8-bits gray-scale image and digital signature | Provides better security | Uses stego-key | [7] |
| 6 | Receiver compatible data hiding | LSB substitution is done only in blue channel | No need of secret key | Only blue channel is used, low capacity | [8] |
| 7 | Pixel indicator technique | In RGB, One is indicator channel and the other two used to hide data | No need of secret key | Indicator channel is not used much | [9] |
| 8 | LSB and PIT | LSB substitution is done on channel chosen by PIT | Randomness of indicator channel | Capacity maybe low | [10] |

The hash based least significant bit 3-3-2 technique is proposed by Koushik Das Gupta, J.K.Mandal and ParamarthaDutta [6] this technique takes eight bits of secret data at a time and put them in LSB of RGB pixel value of the carrier media in 3, 3, 2 order respectively.

## 3. PROPOSED HASH BASED LEAST SIGNIFICANT BIT 2-3-3 TECHNIQUE.

A hash based least significant bit technique is proposed. A color image is considered as a cover media and secret data is embedded in this cover media as payload**.** The proposed technique takes eight bits of secret data at a time and put them in LSB of RGB (Red, Green and Blue) pixel value of the cover image in 2, 3,3 order respectively. Such that out of eight (08) bits of message five (05) bits are inserted in R and G pixel and remaining three (03) bits are inserted in B pixel. The detailed technique has been depicted in Figure 2(a) and (b). An illustration of the same is given in section 4. This distribution pattern is taken because it is giving better results in terms of MSE and PSNR. The proposed method is not tested for the case of compressed images.

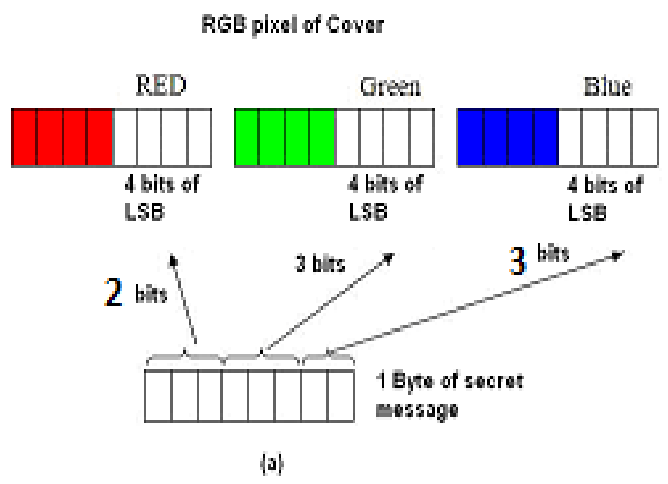

(a)

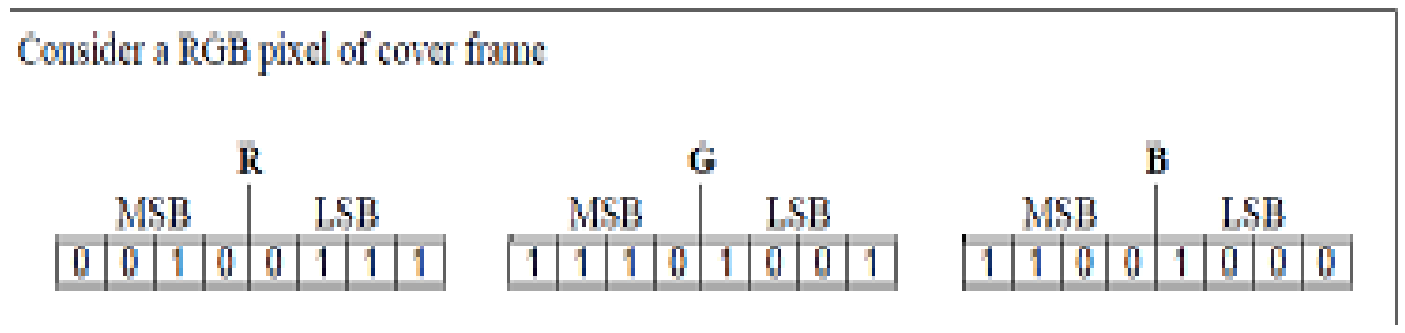

(b)

Figure 2(a) and (b) An example of a cover pixel

Suppose 240 is value of secret image its binary value is 11110101 it is distributed in the order of 2 -3-3 to be embedded in LSB of RGB pixels respectively.

The hash function is as shown below,

$$m = k \,\%\, l \quad (1)$$

where, m is LSB bit position within the pixel, k represents the position of each hidden image pixel and l is number of bits of LSB which is 4 for the present case.

Let the hash function of Equation (1) return values m=1,2 for red, m=3,4,1 for green and m=2, 3, 4 for blue

So, after embedding the secret data in the particular pixel of cover image, The RGB pixel value of the stego image as below

00100111--------------Red
11101110--------------Green
11001011--------------Blue

The embedding positions of the eight bits out of the four (4) available bits of LSB is obtained using a hash function given in equation(1). The bits are distributed randomly using hash function which increases the security of the technique compared to other LSB based techniques [13, 14]. After embedding secret image in the cover image it will become a stego image. The intended user follows the reverse steps to retrieve the secret data. Using the same hash function which is known to the receiver, the data of the secret message is regenerated. The algorithm of the proposed 2-3-3 hash technique has been given in section 4.

# 4. ALGORITHM OF HASH BASED LEAST SIGNIFICANT BIT 2-3-3 TECHNIQUE:

The proposed 2-3-3 algorithm, for encoding and decoding are given in this section. Encoding technique is given in section 4.1 whereas decoding technique is given in section 4.2.

## 4.1. Algorithm of Encoding

Step 1: Input cover image file
Step 2: Read required secret image file to be hidden
Step 3: Take 4 LSB bits of each Red, Green, Blue pixels of the cover image.
Step 4: Obtain the position for inserting the secret data into cover image using hash function given in equation 1.
Step 5: Embed the eight bits of the secret image into 4 bits of LSB of RGB pixels of the cover image in the order of 2,3,3 respectively using the position obtained from step 4.
Step 6: Repeat steps 3 to 5 until all pixels of secret image are embedded in cover image.

## 4.2. Algorithm of Decoding

Step 1: Input stego image file
Step 2: Take 4 LSB bits of each Red, Green, Blue pixels of the stego image.
Step 3: Obtain the position of embedded bits of the secret data using hash function given in equation 1.
Step 4: Retrieve the bits using these positions in the order of 2,3,3 respectively, using the position obtained from step 3.
Step 5: Reconstruct the secret information.
Step 6: Repeat steps 3 to 5 until all pixels of secret image embedded are retrieved.

# 5. RESULTS

Steganography techniques are measured by two attributes, imperceptibility and capacity. Additionally, as an objective measure, the Mean squared Error (MSE), Peak Signal to Noise Ratio (PSNR) and Normalized absolute error (NAE) and Structural similarity index( SSIM) between the stego image and its corresponding cover image are observed. The quantities are given as below, The PSNR is calculated using the equation (2).

$$PSNR = 10\log_{10} L^2/MSE \qquad (2)$$

Where L is peak signal level for an image. The value of MSE is calculated by Equation (3).

$$MSE= \frac{1}{HW} \sum_{i=1}^{H} \sum_{j=1}^{W} (P(i,j) * S(i,j))^2 \qquad (3)$$

Where H and W are height and width and P(i, j) represents the original image and S(i, j) represents corresponding stego image.

Maximum payload (bits per byte/bpb) for the technique has also been obtained i.e. maximumamount of data that can be embedded into the cover image without losing the fidelity of theoriginal image. In the proposed scheme eight bits of data are embedded in 1 pixel of the coverimage.

The **structural similarity** (SSIM) index is a method for measuring the similarity between two images. The SSIM index is a full reference metric; in other words, the measuring of image quality based on an initial cover image before embedding the secret image as reference. SSIM considers image degradation as perceived change in structural information. Structural information is the idea that the pixels have more inter dependencies especially when they are spatially near. The SSIM metric is calculated on various windows of an image. The window can be displaced pixel-by-pixel on the image but the authors prefer to use only a subgroup of the possible windows to reduce the complexity of the calculation.

Normalized absolute error (NAE) computed by Eq.4.is a measure of how far is the stego image from the original cover image with the value of zero being the perfect fit. Big value of NAE indicates poor quality of the resulting image after embedding. The value of NAE is calculated using the equation (4).

$$\text{NAE} = \sum_{i=1}^{H} \sum_{j=1}^{W} \mid P(i,j) - S(i,j) \mid \div \sum_{i=1}^{H} \sum_{j=1}^{W} \mid S(i,j) \mid \qquad (4)$$

The proposed method is evaluated using a set of JPEG images as shown in Table2. The cover images of size 580x580 and 400X400 are used to hide different secret images of size 128X128. The resulting stego images are compared with original cover images to calculate MSE and PSNR values, The proposed 2-3-3 algorithm provides better results compared previous 3-3-2[6] method in terms of MSE and PSNR , NAE, SSIM values. The results of proposed method and previous 3-3-2[6] method are provided in Table 1. A graph is drawn as in figure(3) to show the clear difference between MSE and PSNR values obtained by proposed method and previous[6] method

Table 1: Results

| | | Values obtained for previous method[6] | | | | Values obtained for Proposed 2-3-3 method | | | |
|---|---|---|---|---|---|---|---|---|---|
| Cover Image | Secret Image | MSE | PSNR | NAE | SSIM | MSE | PSNR | NAE | SSIM |
| Pic400.jpg | Lena128.jpg | 11.1738 | 37.6828 | 0.0485 | 0.9672 | 3.7532 | 42.4208 | 0.0080 | 0.9987 |
| Pic400.jpg | Index128.jpg | 9.1791 | 38.5368 | 0.0339 | 0.9869 | 2.8701 | 43.5858 | 0.0052 | 1.0000 |
| Pic580.jpg | Lena128.jpg | 3.4550 | 42.7804 | 0.0070 | 1.0034 | 1.8795 | 45.4243 | 0.0036 | 0.9997 |
| Pic580.jpg | Index128.jpg | 2.2545 | 44.6343 | 0.0045 | 1.0017 | 1.397 | 46.7121 | 0.0023 | 1.0001 |
| Pic580.jpg | Big1124.jpg | 3.2861 | 42.9980 | 0.0066 | 1.0032 | 1.7792 | 45.6625 | 0.0033 | 1.0000 |
| Blueland580.jpg | Lena128.jpg | 10.5965 | 37.9132 | 0.0304 | 1.0431 | 1.8959 | 45.3866 | 0.0036 | 0.9998 |
| Blueland580.jpg | Index128.jpg | 7.4208 | 39.4603 | 0.0196 | 1.0276 | 1.4360 | 46.5931 | 0.0024 | 1.0002 |
| Blueland580.jpg | Big1124.jpg | 10.6427 | 37.8943 | 0.0285 | 1.0404 | 1.7997 | 45.6128 | 0.0034 | 1.0001 |

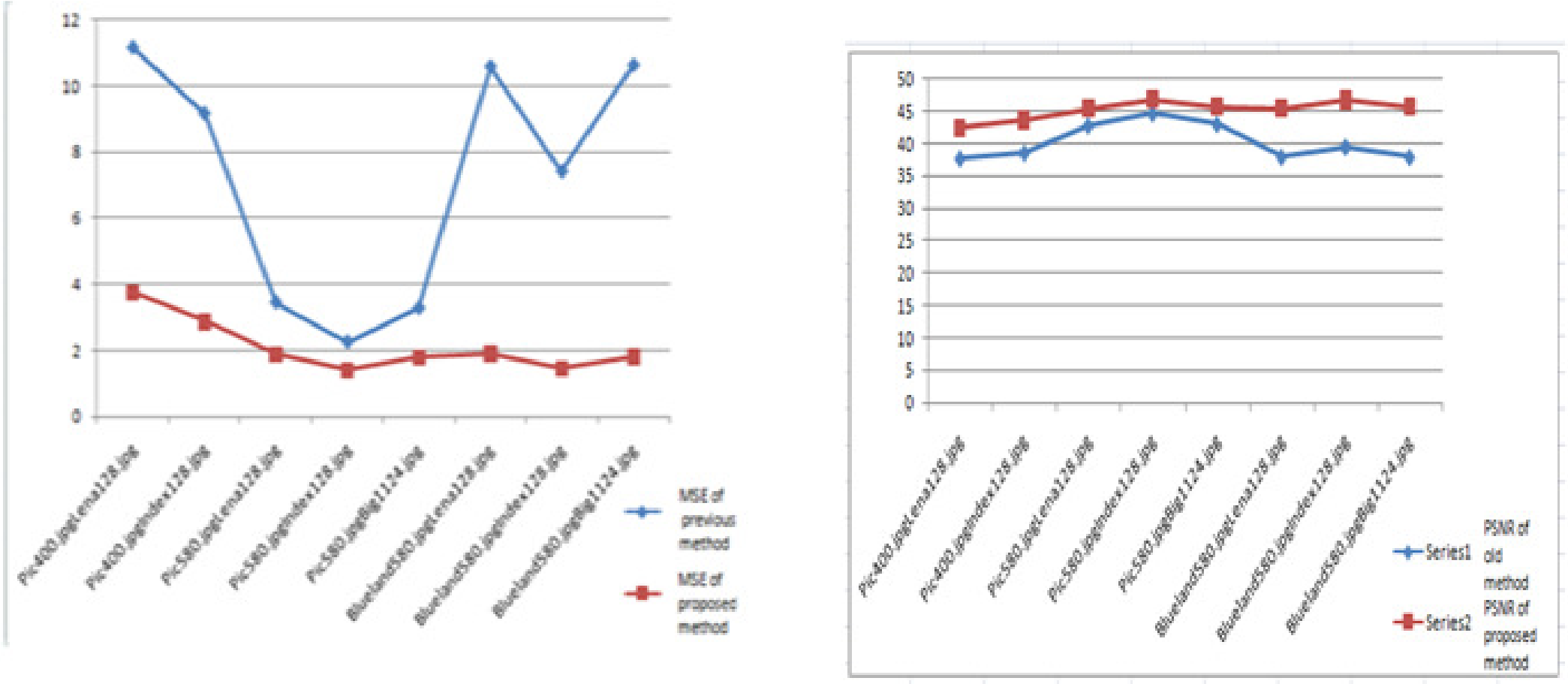

Figure 3  Comparison of MSE and PSNR values

# 6. CONCLUSION

The need for information security is increasing day by day as many people are depending on internet for their daily needs. .An  algorithm HASH BASED LEAST SIGNIFICANT BIT 2-3-3 IMAGE STEGANOGRAPHY is proposed The proposed 2-3-3 algorithm provides better results compared previous 3-3-2 [6] method in terms of MSE and PSNR , NAE, SSIM values. The results of  proposed method and previous [6] method are provided in Table 1.

With a comparison between the proposed algorithm and previous method[6] technique considered by this study, the proposed technique shows promising results as shown in Table 1 and figure 2. There is a drastic improvement in MSE and PSNR values. As an example in case of Cover image pic400.jpg and secret image lena128.jpg, proposed method gets MSE value as 3.7532 and previous method [6] gets MSE value as 11.1738.  So we can conclude that proposed method provides clearly better results. About security enhancing, as a future work Implementing an encryption algorithm for providing more security for secret image can be done.

Table 2 Test bed of Cover and Secret Images along with corresponding stego Images and Recovered Images

| | | | |
|---|---|---|---|
| 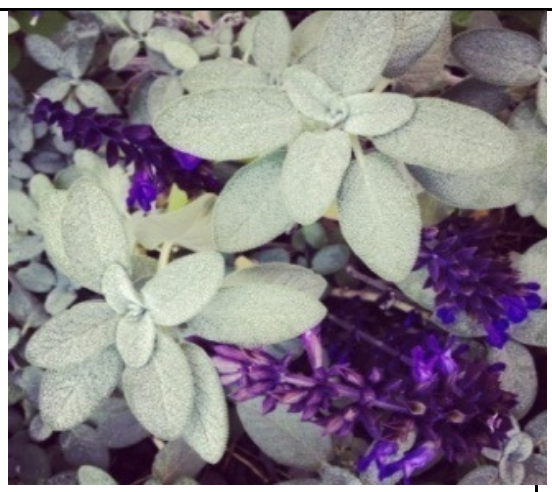<br>CoverPic580.jpg | 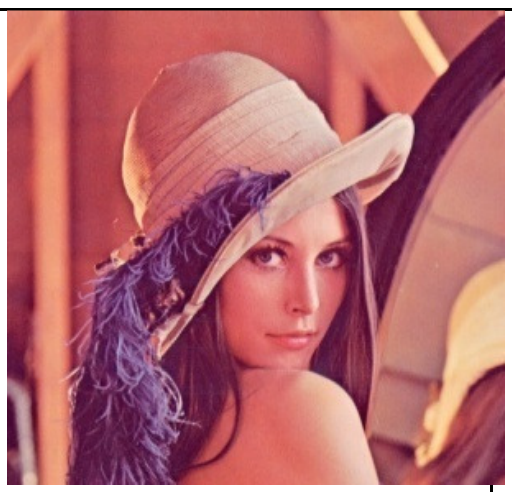<br>Lena.jpg | 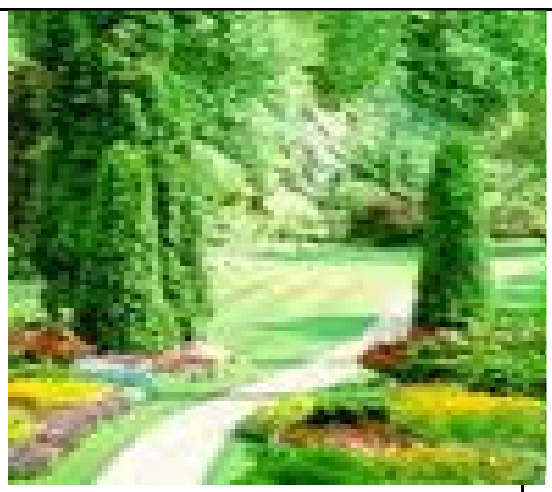<br>Index128.jpg | 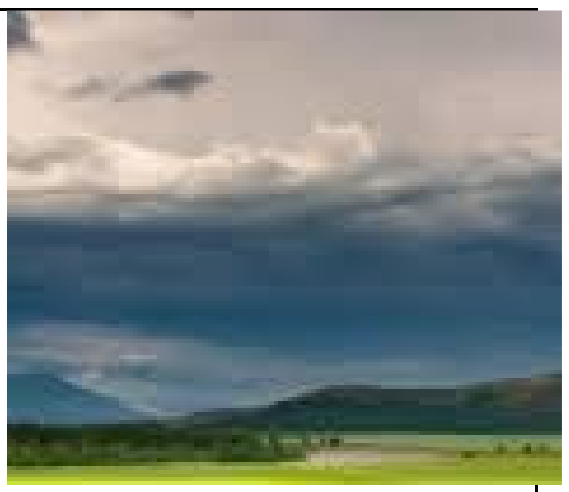<br>Big1124.jpg |
| 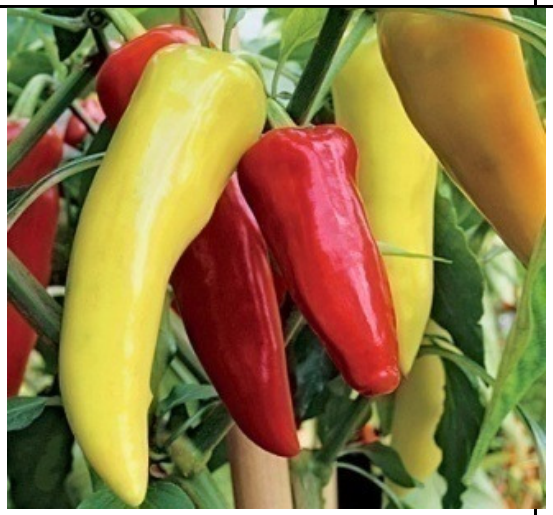<br>pic400.jpg | 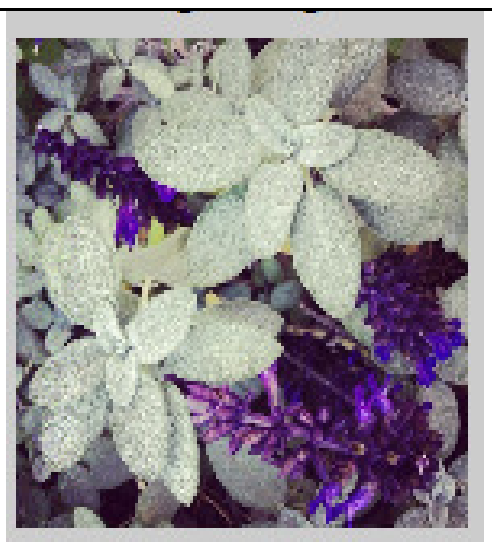<br>StegoLena.jpg | 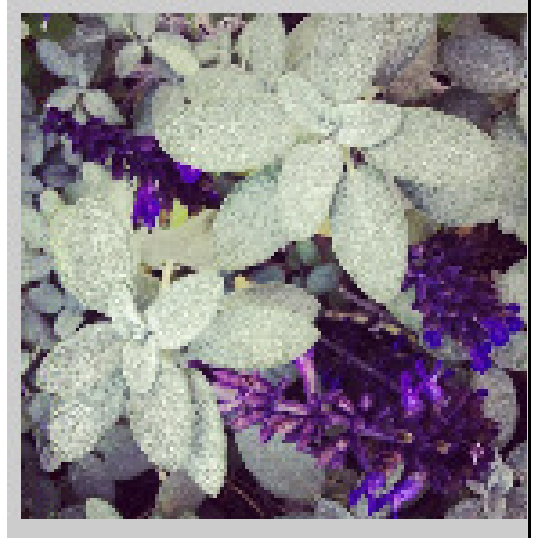<br>StegoIndex128.jpg | 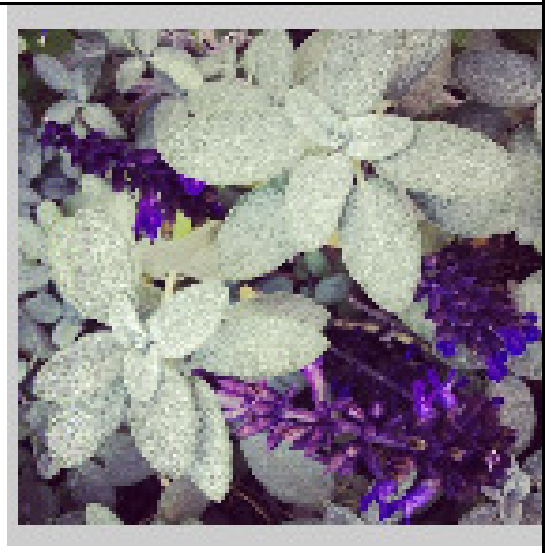<br>StegoBig1124.jpg |
| 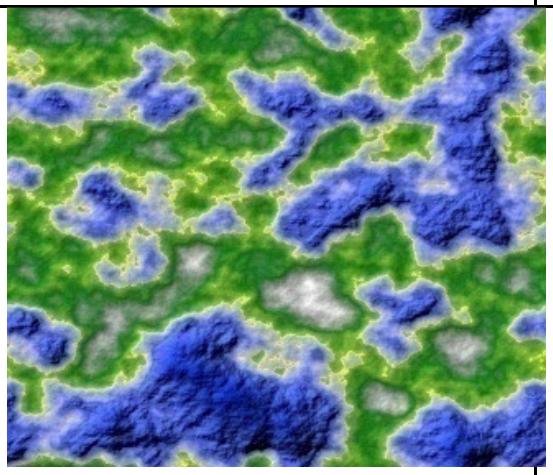<br>CoverBlueland.jpg | 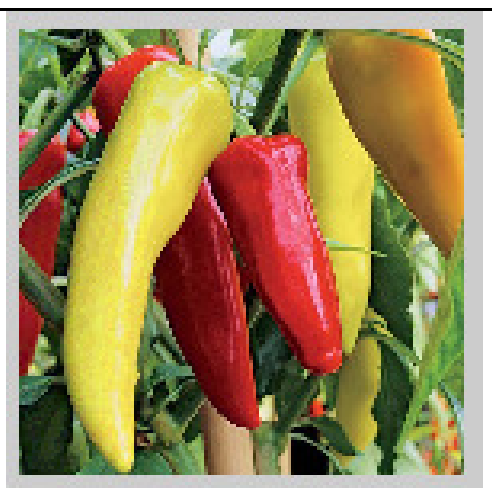<br>StegoLena.jpg<br>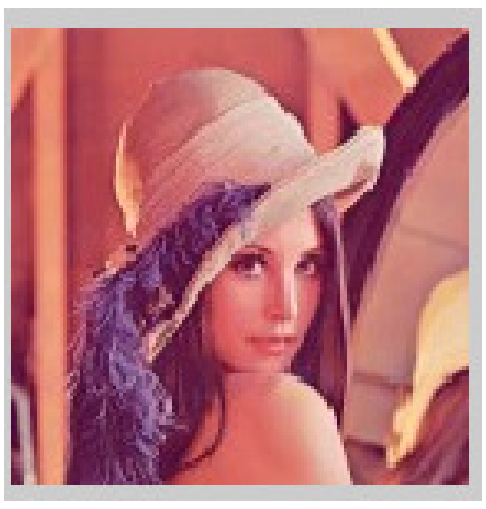<br>Recovered Image | 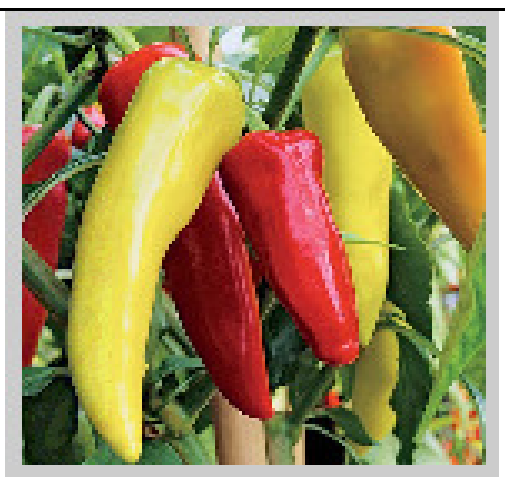<br>StegoIndex128.jpg<br>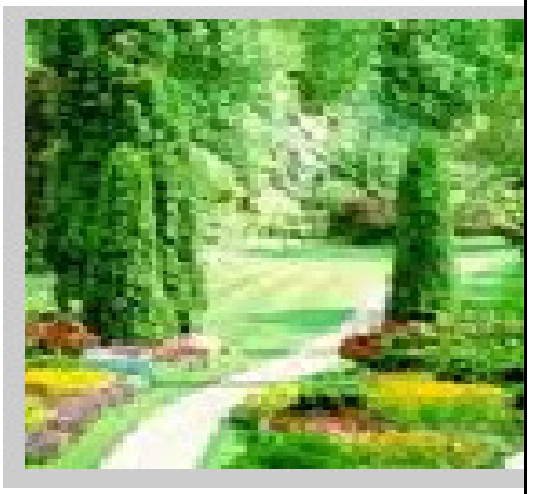<br>Recovered Image | 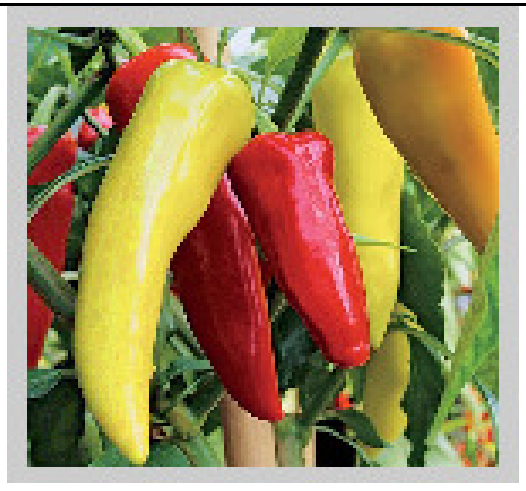<br>StegoBig1124.jpg<br>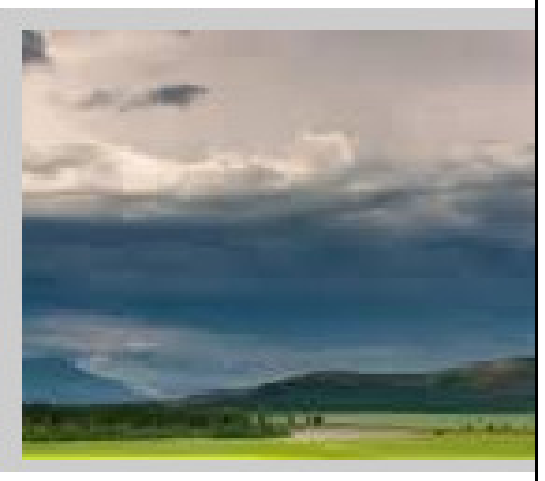<br>Recovered Image |

## 7. ACKNOWLEDGEMENTS

The authors would like to thank principal and staff of JNNCE for their support in this work.

## AUTHORS

**G. R. Manjula**

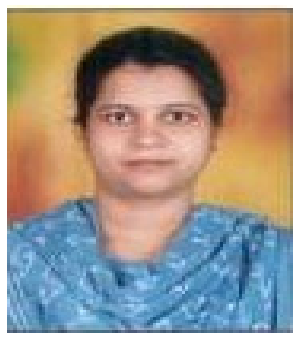

She has obtained M.Tech degree in Networking and Internet Engineering from VTU, Belgaum in 2006 and she is working as Associate Professor in JNN College of Engg. Shimoga, Karntaka, India. She is working for her doctoral degree in Computer Science. Her research areas of interests are Data hiding, Data Embedding, Cryptography and Image Processing.

**Dr. Ajit Danti**

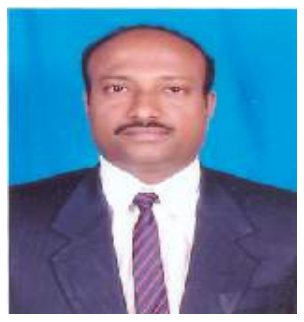

He has obtained Ph.D(Comp.Sc. & Tech) fromGulbarga University, Gulbarga in 2006, Karnataka, INDIA. Currently working as a Professor & Director. Dept. of Computer Applications, Jawaharlal Nehru National College of Engineering, Shimoga, Karnataka, INDIA.His research areas of interests are Computer Vision, Image Processing and Pattern Reco gnition.He has published two books and 30 research papers in peer reviewed International Journals and conferences.